\documentclass{iopart}

\usepackage[dvips]{graphicx}

\def\la{\langle}
\def\ra{\rangle}

\newcommand{\be}{\begin{equation}}
\newcommand{\ee}{\end{equation}}
\newcommand{\ba}{\begin{eqnarray}}
\newcommand{\ea}{\end{eqnarray}} 

\newcommand{\heff}{H_{\rm eff}}

\newcommand{\cpt}{\cal{PT}}

\begin{document}
\date{\today}
\title{Real eigenvalues in non-Hermitian quantum physics}
\author{Ingrid Rotter}
\address{Max-Planck-Institut f\"ur Physik komplexer
Systeme, D-01187 Dresden, Germany }

\begin{abstract}
The dynamics of open quantum systems is determined by avoided and
true crossings of eigenvalue trajectories of a non-Hermitian 
Hamiltonian. The phases of the eigenfunctions are not rigid 
so that environmentally induced  spectroscopic redistribution 
processes may take place and a dynamical phase transition may occur.
Due to the formal equivalence between the quantum mechanical
Schr\"odinger equation and the optical wave equation in 
$\cpt$ symmetric  lattices, the dynamics of the system is determined
also in this case by  avoided and true crossings of eigenvalue 
trajectories of the non-Hermitian Hamiltonian.
In contrast to the eigenvalues characterizing an open quantum system, the
eigenvalues describing the $\cpt$ symmetric optical lattice 
are real as long as the influence
of the environment (lattice) onto the optical wave equation is small. 
In the regime of avoided level crossings, 
the symmetry  is  destroyed, the eigenvalues become complex
and a dynamical phase transition occurs similar as in the open quantum
system.
The redistribution processes in the regime of  avoided level crossings
allow to design systems with desired properties in a broad parameter range.

\end{abstract}


\maketitle

The role of non-Hermitian Hamilton operators in quantum physics is one
of the topical questions studied in recent papers. 
There are two main directions: one goes back to the result obtained by   
Bender and Boettcher \cite{bender} according to which 
a wide class of ${\cal PT}$ symmetric non-Hermitian Hamiltonians 
provides entirely real spectra.
The other (much older) direction  relates the non-Hermiticity of the
Hamiltonian to the fact that quantum systems are embedded into an
environment, see  the recent review  \cite{top}. In this case, 
environmentally induced effects appear, e.g. the finite lifetime of most
states. Also in this theory,
some eigenvalues of the non-Hermitian Hamiltonian are real.
These states, called {\it bound states in the continuum}, are 
related in 1985 \cite{wintgen} to the avoided level crossing phenomenon.  
This relation is confirmed in further studies on realistic systems, see
\cite{top}.
In order to receive a better understanding of non-Hermitian quantum
mechanics, it is necessary to consider the similarities and
differences between these two different directions.
This is important particularly with regard to 
a realistic evaluation of the experimental 
studies suggested in  \cite{ruschhaupt,makris1,makris2,makris3} and
performed in \cite{mumbai}.

Common to both approaches is that the eigenvalues
of non-Hermitian operators may cross in the complex plane. These
crossing points are singular points and called mostly exceptional
points in the mathematical literature \cite{kato}. They are of measure
zero. In open quantum systems, they play however an important role 
due to their non-trivial topological structure \cite{ali}. The eigenvalue
trajectories mostly avoid crossing, and these avoided crossings
influence the properties of the system in a broad parameter range:
a dynamical phase transition takes place \cite{top}. In the present paper, 
it will be shown that the true and avoided level crossings
play an important role also in $\cpt$ symmetric optical lattices.
They cause a dynamical phase transition in a similar manner as in open
quantum systems. In both cases, they are responsible for the appearance and 
disappearance, respectively, of real eigenvalues at high level density.
In order to prove this result, it is suggested to study experimentally
the phase rigidity of the eigenfunctions of the non-Hermitian Hamilton
operator also in $\cpt$ symmetric optical lattices. 

The phenomenon of avoided level crossings is known in
quantum mechanics since many years. In 1932, Landau 
and Zener \cite{landau} published separately the  
analytic solution to the equations of motion governing the transition 
dynamics of a 2-level quantum mechanical system. The two discrete 
states do not cross.  Their smallest distance is reached at
the diabolic point the topological structure of which
is characterized by the Berry phase \cite{berry}.
By analytical continuation into the continuum of scattering wave
functions, the crossing point of the two states can be found 
\cite{ro01,solov}. It is called {\it hidden crossing} in atomic
physics studies \cite{solov}.

An extension of the conventional quantum mechanics of isolated systems 
to  the quantum mechanics of open systems is discussed in many
papers, see the recent review \cite{top}.  In such a treatment, the 
environment (continuum of scattering wave functions) 
is included into the formalism from the very
beginning:  the function space contains not only the discrete states 
of the isolated system but also the continuous scattering wave
functions of the environment. As a consequence, the
Hamilton operator describing the (multi-level) system is non-Hermitian,
\begin{eqnarray}
H_{\rm eff} = H_B + \sum_C V_{BC} \frac{1}{E^+ - H_C} V_{CB} \, . 
\label{heff}
\end{eqnarray}  
Here, $H_B$ is the Hamilton operator describing the closed 
(isolated) system with discrete eigenstates, 
$G_P^{(+)} \equiv (E^+ - H_C)^{-1}  $ is the Green function in the continuum,
$V_{BC},~V_{CB}$ describe the coupling of the closed system to the 
continuum, and $H_C$ is the Hamiltonian describing the
environment of scattering wave functions. The boundary conditions are
taken into account in the two subspaces in a straightforward manner. 
Formally, the second term of
$\heff$ is of second order. However, it becomes the dominant term in the
neighborhood of the singular crossing points \cite{top}.
The operator $H_{\rm eff}$ is  symmetric. 
Its eigenvalues $z_\lambda$ and eigenfunctions $\phi_\lambda$ are complex,
\begin{eqnarray}
(H_{\rm eff} - z_\lambda)\,\phi_\lambda =0 \; .
\label{phi}
\end{eqnarray}
The eigenvalues provide not only the energies 
$E_\lambda$ of the resonance states but also their widths $\Gamma_\lambda$, 
~$z_\lambda = E_\lambda - \frac{i}{2} ~\Gamma_\lambda $. 
The eigenvalues of states below the lowest particle decay threshold are real
(according to $\Gamma_\lambda =0$), while those of states 
above the threshold are complex, generally (according to 
$\Gamma_\lambda \ne 0$). 
The eigenfunctions $\phi_\lambda$  of $H_{\rm eff}$ are  biorthogonal,
and $\langle \psi_\lambda | = \langle \phi_\lambda^{*} |$ because of
the symmetry of $\heff$. The normalization condition 
$\langle\phi_\lambda^*|\phi_{\lambda }\rangle = (\phi_{\lambda })^2 $
fixes only two of the four free parameters \cite{gurosa}. 
This freedom can be used  in order to provide a smooth transition from 
an open quantum system (with, in general, nonvanishing decay widths
$\Gamma_\lambda$ of its states and biorthogonal wave functions) 
to the corresponding closed one (with 
$\Gamma_\lambda \to 0$ and  wave functions  normalized
in the standard manner): $\langle\phi_\lambda^*|\phi_{\lambda }
\rangle \to \langle\phi_\lambda|\phi_{\lambda }\rangle =1 $ 
if $V_{BC}, ~V_{CB} \to 0 $ in (\ref{heff}).
That means, the orthonormality conditions can be chosen as
\begin{eqnarray} 
\langle\psi_\lambda|\phi_{\lambda '}\rangle = \delta_{\lambda, \lambda '} 
\label{biorth1} 
\end{eqnarray}
with the consequence that 
\begin{eqnarray}
A_\lambda  \equiv  \langle\phi_\lambda|\phi_{\lambda}\rangle \ge 1 ~~;
~~~|B_\lambda^{\lambda '}|  \equiv 
|\langle \phi_\lambda | \phi_{\lambda ' \ne \lambda} \rangle| ~\ge ~0  
\label{biorth2b}
\end{eqnarray}
when $\psi_\lambda = \phi_\lambda^*$.
In the regime of overlapping resonances, one has $A_\lambda >1$ and 
$|B_\lambda^{\lambda '}| \ne 0$ due to the second term of (\ref{heff}),
i.e. due to the interaction of the resonance states via the continuum
of scattering wave functions.
The strength of this interaction is given by the non-diagonal matrix
elements of (\ref{heff}). The relation $|B_\lambda^{\lambda '}| \ne
0$ means that the two eigenstates $\lambda$ and $\lambda '$ are 
not orthogonal to one another ({\it skewness of the modes}). 

In this formalism, the conventional quantum mechanics with Hermitian
Hamilton operator appears to be the limiting case with $V\to 0$ where $V$
is the overall interaction between system and environment. The  hidden
crossings become {\it true crossing points} of the complex eigenvalues. 
They are singular points and 
their topological structure differs from that of a diabolic point: the
geometric phase is larger than the Berry phase by a factor two
\cite{gurosa}, as proven
experimentally in a study  on a microwave billiard \cite{demb1}.

The crossing points of two complex eigenvalues can be studied best by
considering the $2\times 2$ symmetrical non-Hermitian Hamiltonian
\begin{eqnarray}
\label{ham2}
H=\left(
\begin{array}{cc}
\epsilon_1 &  \omega \\ \omega & \epsilon_2
\end{array}
\right)
\end{eqnarray}
where $\epsilon_\lambda$ are complex and stand for the energies and
widths of two isolated resonance states, $\lambda =1, ~2$. The
interaction $\omega$ between the two states is complex, generally.
Both, the $\epsilon_\lambda$ and the $\omega$, may depend on a
parameter $X$ by means of which the system  can be 
controlled. The eigenvalues of (\ref{ham2}) are   
\begin{eqnarray}
\label{eig2}
E_\pm &=&
\frac{\epsilon_1 + \epsilon_2}{2}
 \pm  \frac{1}{2}\sqrt{(\epsilon_1-\epsilon_2)^2 +4\omega^2} 
\, .
\end{eqnarray}
The two trajectories $E_+(X)$ and $E_-(X)$ cross when 
$(\epsilon_1-\epsilon_2)/(2\omega) = \pm ~i $. This
condition can not be fulfilled for discrete states since $\epsilon_\lambda$
and $\omega$ are real in this case.
The  eigenfunctions of $ H$ at the crossing point are 
linearly dependent 
\begin{eqnarray}
\phi_+^{cr} \to \, \pm\, i\, \phi_-^{cr} \; \; ;
\qquad  \phi_{-}^{cr} \to \, \mp \, i \, \phi_+^{cr} \; .
\label{r1a} 
\end{eqnarray}
At the crossing point, the phase  of the wave function jumps by $\pi /4$.
Here, $A_\lambda \to \infty, ~|B_\lambda^{\lambda '}| \to \infty$. 
When the two states avoid crossing,  $A_\lambda$ and $|B_\lambda^{\lambda '}|$
are finite (but different from 1 and 0, respectively) 
and the two states are mixed in a certain finite range of
the parameter considered. This range is the larger the smaller
the decay widths of the states are. It is especially large
when discrete states avoid crossing \cite{ro01}.

The normalization condition (\ref{biorth1}) entails that 
the phases of the eigenfunctions of $\heff$ in the overlapping regime
are not rigid: the normalization condition
$\la\phi_\lambda^*|\phi_{\lambda}\ra =1$ is fulfilled, in this regime, 
only when Im$\langle \phi_\lambda^*|\phi_\lambda\rangle \propto $
Re$~\phi_\lambda \cdot$  Im$~\phi_\lambda =0$, i.e. 
by rotating the wave function through a certain angle $\beta_\lambda$. 
The phase rigidity defined by
\begin{eqnarray}
r_\lambda = 
\frac{\langle \psi_\lambda| \phi_\lambda \rangle }
{\langle\phi_\lambda | \phi_\lambda \rangle} = 
\frac{1}{({\rm Re}\, \phi_\lambda) ^2 + ({\rm Im}\, \phi_\lambda) ^2}= 
\frac{1}{A_\lambda}
\label{ph2} 
\end{eqnarray}
is a useful measure of the rotation angle $\beta_\lambda$. 
When the resonance states are distant from one another and the 
eigenfunctions are (almost) orthogonal in the standard manner, it is 
$r_\lambda \approx 1$ due to $ \langle\phi_\lambda |\phi_\lambda\rangle$ 
$\approx \langle\phi_\lambda^*|\phi_\lambda\rangle$. 
In approaching a crossing point in the complex energy plane, 
we have $A_\lambda \to \infty$ and the two wave functions become linearly
dependent, see (\ref{r1a}). Therefore $1\ge r_\lambda \ge 0$.
This result is proven experimentally in a study on a microwave cavity 
\cite{demb2} (see \cite{top} for the interpretation). Since the phases 
are not rigid in the regime of avoided level crossings, some of the
eigenfunctions may align to the scattering wave functions of the
environment. By this, spectroscopic redistribution processes 
become possible which cause a dynamical phase transition \cite{top}. 

The most interesting difference between the Hermitian and
non-Hermitian quantum physics is surely the fact that the phases of the 
eigenfunctions of the Hamiltonian are rigid in the first case,
while they are variable in the second case. By this, the non-Hermitian 
quantum physics is able to describe environmentally induced effects
(feedback between system and environment), especially dynamical phase
transitions.  These effects, visible in spectroscopic studies, 
occur in the regime of avoided (and true) level crossings 
where $r_\lambda < 1$. 

As shown in \cite{ruschhaupt,makris1,makris2,makris3}, 
the quantum mechanical Schr\"odinger equation and the optical wave
equation in $\cpt$ symmetric optical lattices are formally equivalent. 
Complex $\cpt$ symmetric structures can be realized 
by involving symmetric index
guiding and an antisymmetric gain/loss profile. The main difference of
these optical systems to open quantum systems consists in the asymmetry of
gain and loss in the first case while the states 
of an open quantum system can only decay (Im$(\epsilon_{1,2}) < 0$
for all states).  Thus, the  modes involved in the non-Hermitian 
Hamiltonian in optics appear in complex conjugate pairs 
while this is not the case in an open quantum system.
As a consequence, the Hamiltonian for $\cpt$ symmetric structures in 
optical lattices may have real eigenvalues when gain and loss are small.
The $2\times 2$ non-Hermitian Hamiltonian may be written as
\begin{eqnarray}
\label{ham2o}
H_o=\left(
\begin{array}{cc}
a &  b \\ c & d
\end{array}
\right), 
\end{eqnarray}
where 
Re$(a) =$ Re$(d) = \varepsilon $ stand for the energies  of the two
modes, Im$(a) = - $ Im$(d) \equiv - \gamma$ describes  gain and loss,
respectively,
and the complex coupling coefficients $b$ and $c=b^*$ stand for the
coupling of the two modes via the lattice. The eigenvalues differ from 
(\ref{eig2}),
\begin{eqnarray}
\label{eig2o1}
E^o_\pm &=& \varepsilon \pm \frac{1}{2}\sqrt{4|b|^2 - \gamma^2}
\, .
\end{eqnarray}
Since $\varepsilon$ and $\gamma$ are real, the $E^o_\pm$ are real when
$4|b|^2 > \gamma^2$ and may cross when $4|b|^2 =
\gamma^2$. When the optical lattices
are studied with vanishing gain term,
it is Im$(a)=0$, Im$(d) = \gamma$ and 
$\tilde E^o_\pm = \varepsilon - \frac{i}{2} \gamma
\pm \frac{1}{2}\sqrt{4|b|^2 - \frac{1}{4}\gamma^2}$.
Therefore, the $ \tilde E^o_\pm + \frac{i}{2} \gamma$
are real in this case when 
$4|b|^2 > \frac{1}{4}\gamma^2$ and may cross when $4|b|^2 =
\frac{1}{4}\gamma^2$. Hence, the energy spectrum depends linearly on
$\gamma$ under these conditions.

The crossing points of eigenvalues are discussed in the mathematical 
literature for many years \cite{kato}. Here they are called mostly {\it 
exceptional points}. They are singular points.
The most important property of these points is that not only two
eigenvalues coalesce  but also the
corresponding eigenfunctions are linearly dependent, see e.g. (\ref{r1a}).
The crossing points are meaningful 
not only for the dynamics of open quantum systems (as discussed above) 
but also for $\cpt$ symmetric optical lattices involving the $2\times 2$ 
non-Hermitian Hamiltonian $H_o$. Also in this case, the
wave functions of the two  states (eigenfunctions of the 
non-Hermitian Hamiltonian $H_o$) are orthogonal to one another
far from the crossing point. However, the
orthogonality is lost in approaching the crossing point.
The eigenmodes  can, generally, be normalized according to  
(\ref{biorth1}) where $ \psi_\lambda$ denotes the left eigenmode.
Using this normalization, $\langle\psi_\lambda|\phi_{\lambda
  '}\rangle \to \langle \phi_\lambda | \phi_{\lambda '}\rangle $ 
with $\gamma \to 0$  (limiting case of vanishing loss [and gain]). 
As in the case of an open quantum system, the phase rigidity 
$r_\lambda$ varies between 1 (limiting case $\gamma \to 0$)
and 0 (limiting case $4|b|^2=\gamma^2$ or $4|b|^2 = \frac{1}{4}\gamma^2$). 
As a consequence, the phases of the eigenmodes of the
non-Hermitian Hamiltonian (\ref{ham2o}) are not  rigid, and  
redistribution processes occur generally  
in the regime of  avoided level crossings.

According to these results one expects some analogy 
between characteristic features of an open quantum system and 
the optical transparency of a $\cpt$ symmetric lattice.
In both cases, the system evolves according to expectations of the
standard Hermitian quantum physics as long as the influence from the
environment (expressed by $\omega$ and $b$, respectively)
is small. In open multi-level systems, the widths of the resonance 
states increase first with increasing coupling $\omega$ between
system and environment, i.e. with increasing degree of resonance overlapping
\cite{top}. However, with further increasing
coupling strength between system and environment,  
the evolution of the open quantum 
system becomes counterintuitive. This has been observed first in
nuclear structure calculations \cite{kleiro} more than 20 years ago. 
In the regime of overlapping resonances (with many avoided level crossings),
most states of the system decouple finally, to a large extent, from the
environment of scattering wave functions while a few states align to
the scattering wave functions and become short-lived. 
This phenomenon called {\it resonance trapping}, is proven
experimentally in a study on a microwave cavity \cite{stm}.
It occurs due to width bifurcation in a hierarchical manner \cite{top},
i.e. one resonance state aligns with a scattering state
by trapping  step by step  other resonance states.
Resonance trapping is   {\it  environmentally induced}:  
a few states become short-lived (i.e. they align with the scattering states)
at the cost of the remaining ones. As a consequence,
transmission through a small system is enhanced when the resonances
overlap (and many  levels avoid crossing), see \cite{top,burosa}. 

A similar behavior is expected in $\cpt$ symmetric optical lattices.
At small loss, the transparency of the system is expected to decrease
with increasing loss. Due to avoided level
crossings, however, the perturbance of the modes by the environment
(lattice) will cause a bifurcation between gain and loss, 
and the eigenvalues will become complex. Hence, 
the transparency will be enhanced with further increasing loss -- in
analogy to the counterintuitive behavior of open quantum systems at
strong coupling between system and environment. 

Important for the physics of the system
is the square root in Eqs. (\ref{eig2}) and 
(\ref{eig2o1}) since it determines the positions of the crossing 
points. In the open quantum system, the imaginary part of the
eigenvalues is always negative corresponding to the fact that the
states of an open quantum system can only decay (time asymmetry
of an open quantum system). Resonance states with vanishing decay width 
(i.e. with real eigenvalues) can appear therefore only due
to spectroscopic redistribution processes (width bifurcation)  
occurring in the regime of overlapping resonances with many 
avoided level crossings. In $\cpt$ symmetric optical lattices, however,
the imaginary part of the eigenvalues may be positive or negative. 
Hence, the Hamiltonian may have real eigenvalues 
as long as $\gamma$ is small (in comparison to $|b|^2$). 
Redistribution processes under the influence of avoided (and true)
level crossings will destroy the symmetry
with the consequence that the eigenvalues of the Hamiltonian will 
cease being real. 

In both cases, the  redistribution processes cause a dynamical phase
transition which does {\it not} 
start at a singular point at which two eigenvalues coalesce.   
They start when the rigidity of the phases 
of the eigenfunctions of the Hamiltonian gets reduced
(i.e. in the regime of avoided level crossings)
so that the interference picture changes. At a certain 
critical parameter value, the redistribution process is completed
in a multi-level system:
a few states are short-lived while most states are trapped. 

As shown in numerical studies, transmission through small open
quantum systems is enhanced (and accelerated) {\it below} the critical
parameter value, indeed. It is correlated with a reduced phase rigidity
\cite{burosa}. However, width bifurcation becomes visible only 
at a parameter value {\it beyond} the critical one. Here it may create
bound states in the continuum, when the system is  spatial symmetric
\cite{top}. In this case, the trapped state has a
vanishing decay width  and is decoupled completely
from the continuum of scattering wave functions. 
This behavior results from the interplay between internal
interaction [involved in $H_B$] and external interaction [described by
the second term of $\heff$ in (\ref{heff})] of the states. It is found 
in numerical calculations for different realistic systems \cite{top}. 

In $\cpt$ symmetric optical lattices,
the  redistribution processes destroying the
symmetry of the system, start also at a parameter value below
the critical one. The  redistribution resulting in
complex eigenvalues of the Hamiltonian, is expected to
become visible therefore {\it before} the critical parameter value
is reached. The redistribution processes  influence significantly
the transparency of  $\cpt$ symmetric lattices. 
Increasing losses at the idler frequency lead to broadband
transparency or amplification at the signal frequency. This result
agrees with the conclusion drawn  in \cite{popov}
by using another formalism.  It would be very interesting to
control experimentally the angle between the two eigenmodes 
in approaching the crossing point. The variation of this
angle  will prove, on the one hand,  one of the
most interesting features of non-Hermitian quantum theory,
namely the non-rigidity of the phases  of the eigenfunctions of the 
Hamiltonian. On the other hand, it will directly
prove to which extent the reduced phase rigidity
in the regime of avoided level crossings is responsible for the
surprising redistribution processes in $\cpt$ symmetric optical lattices.

Finally, it should be remarked that environmentally induced spectroscopic
redistribution processes (dynamical phase transitions)
have been observed experimentally in 
different studies. Phase lapses are observed in the transmission 
through small quantum dots \cite{heiblum} what could not be explained
in the framework of Hermitian quantum mechanics in spite of much effort.
An explanation  by means of the non-rigid phases of the eigenfunctions of the
non-Hermitian Hamiltonian is possible \cite{murophas,top}. 
Well known for many years are the narrow densely lying compound
nucleus resonances in heavy nuclei which differ fundamentally from the
few much broader resonances in light nuclei. 
Another example for a dynamical phase transition is studied
experimentally and theoretically in quantum chemistry \cite{pastawski}. 
It is analyzed by using a Keldish-like formalism. 

Summarizing it can be stated the following. 
The  non-Hermitian Hamiltonian allows to describe
environmentally induced phenomena in quantum systems 
and in systems being formally equivalent to them,
in a straightforward manner. These effects are important in the regime of
overlapping resonances where the dynamics of the system is
determined by avoided (and true) level crossings.
Here,  a dynamical phase transition takes place
and the conditions for the appearance and disappearance of real
eigenvalues change.
In contrast to the general statements for open quantum systems, 
real spectra of $\cpt$ symmetric non-Hermitian 
Hamiltonians can be realized in optics at small coupling between
system and environment (lattice) due to the asymmetry between
gain and loss. The eigenvalues are real under this condition and become 
complex only in the scenario with avoided (and true) level crossings.
According to the results obtained in the present paper, 
the dynamical phase transition in optical lattices is
related to the reduced phase rigidity of the eigenfunctions of the 
non-Hermitian Hamiltonian in the regime of avoided level crossings.
It causes  $\cpt$ symmetry breaking.

It would be highly interesting to study experimentally
the correlation between phase rigidity and dynamical phase transition
also in optical lattices. The results are expected to be similar to 
those observed experimentally in an open microwave cavity 
in which the crossing point of two eigenvalue trajectories is 
approached by varying a parameter \cite{demb2} (see \cite{top} for the
interpretation of the experimental data).  
They are expected to show also features being similar to those
obtained theoretically for the phase transition in a toy model
\cite{jung} or for the transmission through quantum dots
in the regime of overlapping resonances \cite{burosa}. 
A proper account of environmentally induced effects 
in non-Hermitian quantum physics allows 
to design systems with  desired properties in a broad parameter range.
\\

\end{document}